\documentclass[aps,pre,amsmath,preprint,showpacs]{revtex4}
\usepackage{graphics,epsfig}
\usepackage{bm}

\begin{document}

\title{Pitchfork and Hopf bifurcation thresholds in stochastic equations
with delayed feedback}
\author{Mathieu Gaudreault, Fran\c{c}oise L\'{e}pine, and Jorge Vi\~nals}
\affiliation{Department of Physics, McGill University, Montreal, Quebec, 
Canada, H3A 2T8.}

\date{\today}

\begin{abstract}
The bifurcation diagram of a model stochastic differential equation with 
delayed feedback is presented. We are motivated by recent research on
stochastic effects in models of transcriptional gene regulation. We start 
from the normal form for a pitchfork bifurcation, and add multiplicative or 
parametric noise and linear delayed feedback. The latter is sufficient to 
originate a Hopf bifurcation in that region of parameters in which there 
is a sufficiently strong negative feedback. We find a sharp 
bifurcation in parameter space, and define the threshold as the point in 
which the stationary distribution function $p(x)$ changes from a delta 
function at the trivial state $x=0$ to $p(x) \sim x^{\alpha}$ at small 
$x$ (with $\alpha = -1$ exactly at threshold). We find that the bifurcation 
threshold is shifted by fluctuations relative to the deterministic limit 
by an amount that scales linearly with the noise intensity. Analytic 
calculations of the bifurcation threshold are also presented in the 
limit of small delay $\tau \rightarrow 0$ that compare quite favorably 
with the numerical solutions even for $\tau = 1$.
\end{abstract}


\maketitle

\section{Introduction}

We study the bifurcation diagram of a nonlinear stochastic differential 
equation that includes delayed feedback. The model equation considered exhibits both pitchfork and Hopf bifurcations. The bifurcation thresholds are obtained as a function of the model parameters, and our results contrasted with two related limits: The deterministic limit of a differential delay equation, and the stochastic bifurcation of the same model without delay.
 
The study of differential delay equations \cite{re:driver77} is an important topic in applied mathematics, with widespread applications in Physics (lasers, liquid crystals), control systems in Physiology (neural and cardiac tissue activity) \cite{re:mackey77,re:glass88}, and Economy (agricultural commodity prices) \cite{re:belair89}. Recent interest has arisen in the mathematical modeling of cellular function at the molecular level, especially in transcriptional gene regulation \cite{re:hasty01}. Feedback regulation is a common motif in cellular networks, with delays arising from the complexity of the underlying network, or from the wide disparity in time scales of the many chemical processes involved in regulation \cite{re:glass88}. For example, DNA is transcribed at a rate of 10 to 100 nucleotides per second, and it may take a delay of the order of minutes  before the transcription factor appears as a finished product in the cell, and hence is available for regulation. Significant delays can also be attributable to the time required for the diffusion of proteins through membranes, so that, for example, the auto regulated feedback on protein production at time $t$ is often proportional to protein concentration at time $t-\tau$, where $\tau$ is known as the delay time. For short delay times, a reaction may be approximated as being instantaneous, and the system as being in quasi equilibrium. However, when the delay is comparable to the characteristic time scales of reaction, the non instantaneous nature of the interactions can no longer be ignored, and delay terms need to be included in the governing equations for the network under study \cite{re:mackey95,re:bratsun05}.

Experimental evidence has been mounting that highlights the importance of stochastic effects in transcriptional regulation \cite{re:kepler01,re:elowitz02,re:bratsun05,re:maheshri07}, not only for natural networks, but for engineered gene circuits and networks as
well \cite{re:elowitz00,re:hasty00}. However, despite the wealth of evidence pointing to the importance of stochasticity in feedback regulation, delays in stochastic models of metabolic feedback are very often neglected, possibly because the resulting stochastic equations are no longer Markovian, and hence rarely tractable analytically. Exceptions include the derivation of a two time Fokker-Planck equation and the study of its small delay time limit in  \cite{re:guillouzic99}, and results on the bifurcation of the first and second moments of a linear stochastic equation with delay \cite{re:mackey95,re:lei07}. We extend these latter results to the analysis of the stationary probability distribution function of a nonlinear model, and discuss in detail the stability of the solutions that results from the interplay of delay and stochasticity.

We focus here on the normal form for a pitchfork bifurcation augmented with multiplicative or parametric noise (additive noise has been shown to have no effect in the case of a delayed equation \cite{re:mackey95}), and linear delayed feedback. The latter is sufficient to originate a Hopf bifurcation in some region of parameters (sufficiently strong negative feedback). Our model for the dynamical variable $x(t)$ is
\begin{equation}
\dot{x}(t) = ax(t) + bx(t-\tau) - x^3(t) + \xi(t)x(t)\; ,
\label{eq:StochmultDDnonlin_ch3}
\end{equation}
where the constant $a$ plays the role of the control parameter, $b$ is the intensity of a feedback loop of delay $\tau$, and $\xi(t)$ is a white, Gaussian noise of intensity $D$. The initial condition is a function $\phi(t)$ specified on $t=[-\tau,0]$. We generally focus on the stationary probability distribution function $p(x)$ for a range of values of $a, b$ and $D$.

The bifurcation diagram for the differential delay equation resulting from Eq. (\ref{eq:StochmultDDnonlin_ch3}) with $\xi = 0$ is known (see, e.g. \cite{re:mackey95}). Linearization around $x=0$ shows that trajectories decay asymptotically to zero according to
\begin{eqnarray}
b  < & -a  & \mbox{ if } b\tau > -1  \; , \label{eq:det_boundary_pitch} \\
\tau\sqrt{b^2 - a^2} < & \cos^{-1}(-\frac{a}{b}) & \mbox{ if } b\tau < -1 \; . \label{eq:det_boundary_hopf}  
\end{eqnarray}
The boundary separating exponentially decaying solutions from exponentially growing solutions is shown as the solid line in Fig. (\ref{fig:bif_diagram}). The upper branch corresponds to a pitchfork bifurcation (real eigenvalue, Eq. (\ref{eq:det_boundary_pitch})), whereas the lower branch corresponds to a Hopf bifurcation (complex eigenvalue, Eq. (\ref{eq:det_boundary_hopf})). In both cases, we show in the figure the case of  $\tau = 1$. The cusp at the intersection of both boundaries is located at $(a,b) = (1/\tau,-1/\tau)$.

The bifurcation threshold of Eq. (\ref{eq:StochmultDDnonlin_ch3}) without 
delay ($b = 0$) is also known, and has been given in \cite{re:graham82b,re:drolet98,re:sanmiguel99}. Recall that an analysis of the linearized equation leads to the unphysical conclusion that the bifurcation threshold depends on the order of the statistical moment of $x(t)$ considered. On the other hand, with a saturating nonlinearity in Eq. (\ref{eq:StochmultDDnonlin_ch3}), a stationary probability distribution of $x$ exists both below and above threshold, thus allowing a proper determination of the bifurcation point. The stationary distribution function of $x$ with $b = 0$ has been found to be \cite{re:graham82b}
\begin{eqnarray}
\alpha \leq -1 & & p_{0}(x) = \delta(x) \; , \label{eq:delta_function} \\
\alpha > -1 & & p_{0}(x) = N x^{\alpha}e^{-\frac{x^2}{2D}} \; ,\label{eq:nlin_prob}
\end{eqnarray}
where the exponent $\alpha = a/D - 1$, and $N$ is a normalization constant. The solution (\ref{eq:nlin_prob}) exists but is not normalizable for $\alpha < -1$ and hence it is not a physically admissible solution. Therefore the stochastic bifurcation threshold is located at $a_{c} = 0$, point at which $p(x)$ changes from a delta function at the the origin to a power law at small $x$ with an exponential cut off at large $x$. Contrary to what an analysis of the moments of $x(t)$ would indicate, the existence of parametric fluctuations has no effect on the location of the bifurcation threshold: both deterministic and stochastic equations exhibit a pitchfork bifurcation at $a_{c} = 0$. We further note that in $-1 < \alpha < 0$, $p(x)$ is unimodal with a singularity at $x=0$, whereas for $\alpha > 0$ the distribution
is bimodal reflecting nonlinear saturation of $x$.

\section{Stochastic bifurcation}

We now turn to the case of delay, $b \neq 0$. Analytical results for the stability of the trivial solution $x=0$ for the {\em linearization} of Eq. (\ref{eq:StochmultDDnonlin_ch3}) have been given in \cite{re:mackey95,re:lei07}, and are shown in Fig. (\ref{fig:bif_diagram}). The bifurcation threshold of the first moment $\langle x \rangle$ is shifted relative to the deterministic delay equation result; only bounds can be given for the stability of the second moment $\langle x^{2} \rangle$ \cite{re:lei07}, and no results are available for $p(x)$. Given the anomalous behavior described above for the stochastic bifurcation of the linear equation with $b = 0 $, it is of interest to determine $p(x)$ for the full model of Eq. (\ref{eq:StochmultDDnonlin_ch3}). Unfortunately, the non Markovian character of this equation has precluded progress along these lines \cite{re:guillouzic99}. 

We have first extended an existing second order algorithm for the integration of stochastic differential equations \cite{re:fox91} to the case of delay. The algorithm needs to take into account trajectories into the past for an interval $\tau$, and also new contributions from the stochastic terms that result from the coupling to the delayed feedback. Derivation of the algorithm is presented in appendix \ref{se:appA}. In the numerical calculations to be presented below, the initial 
condition is a constant function in $[-\tau,0]$ for each trajectory, with the 
constant being drawn from a Gaussian distribution of zero mean
and variance $1$. The time step used in the numerical integration is $\Delta t = 0.01$. We also present approximate analytic calculations in the limit of small delay time $\tau \rightarrow 0$, following the approach of Frank \cite{re:frank05}. Reasonable agreement is found with our numerical calculations for $\tau = 1$.

\subsection{Pitchfork bifurcation}

\begin{figure}[h]
\includegraphics[width=6in,clip]{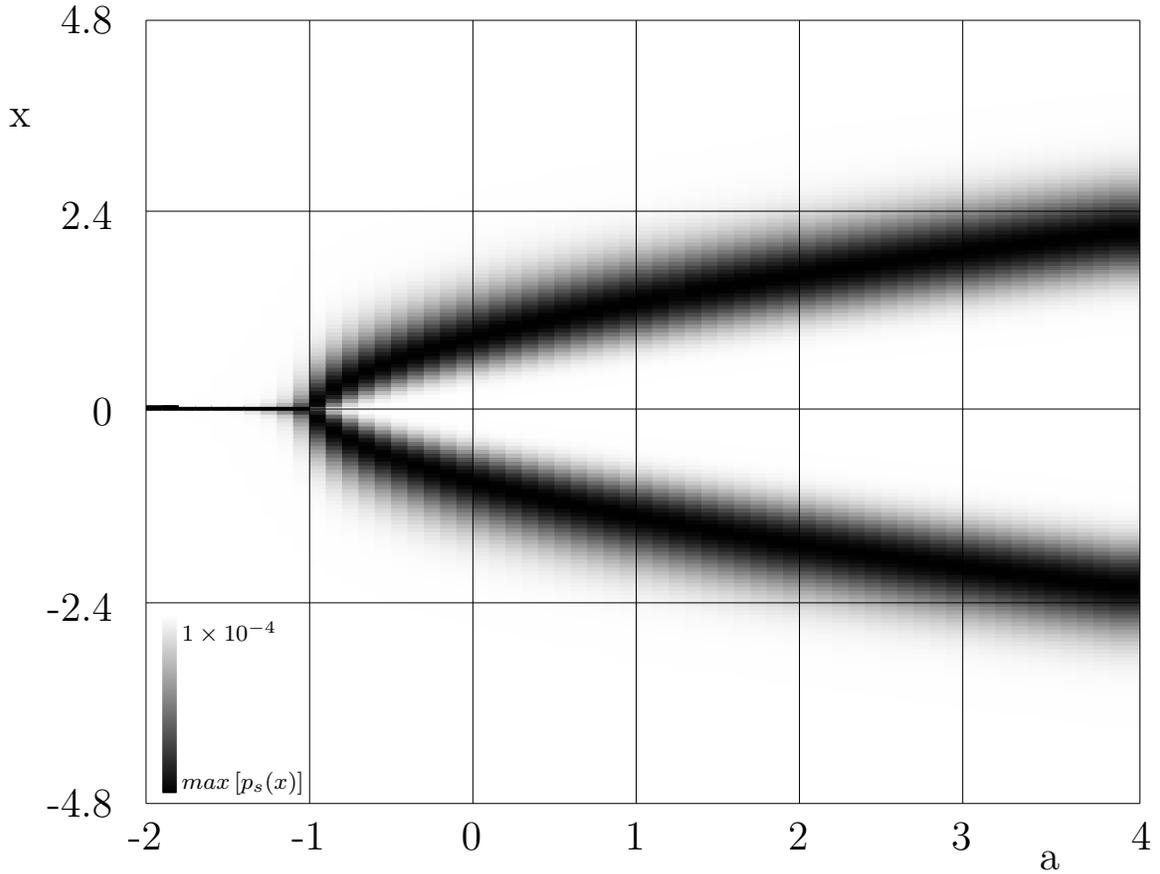}
\caption{Long time histogram of $x$ (in grey scale) as a function of the control parameter $a$ with $b = 1$, $\tau = 1$, and $D = 0.3$.  The histograms have been collected in the time interval $t \in (50, 80)$ and further averaged over $1 \times 10^{6} $ independent runs. In the absence of noise, the critical value of the control parameter for instability is $a_{c} = -1$. We find instead that the bifurcation from a delta function to a power law distribution occurs at $a_{c} = \simeq -1.18$ instead for this set of parameters. Fluctuations around $x = 0$ are observed for $a < a_{c}$ due to the finite length of the time series.}
\label{fig:histogram}
\end{figure}

A qualitative view of the bifurcation of Eq. (\ref{eq:StochmultDDnonlin_ch3}) is given in Fig. (\ref{fig:histogram}). Equation (\ref{eq:StochmultDDnonlin_ch3}) has been integrated numerically, and histograms of $x(t)$ computed once trajectories approach a statistical steady state. The histograms shown correspond to $10^{6}$ independent trajectories for each value of $a$. For $a \lesssim -1$, the histogram is approximately a delta function at $x=0$. As discussed further below, we observe long transients in $x(t)$ until it eventually decays to $x=0$. At a critical value $a_{c} \approx -1$, the bifurcation point, a broad distribution emerges, although the most likely value remains $x = 0$. At larger values of $a$, the histogram becomes bimodal. The histogram shown in the figure corresponds to the direct bifurcation branch, but a qualitatively similar graph is obtained for the Hopf bifurcation.

\begin{figure}[h]
\includegraphics[width=6in,clip]{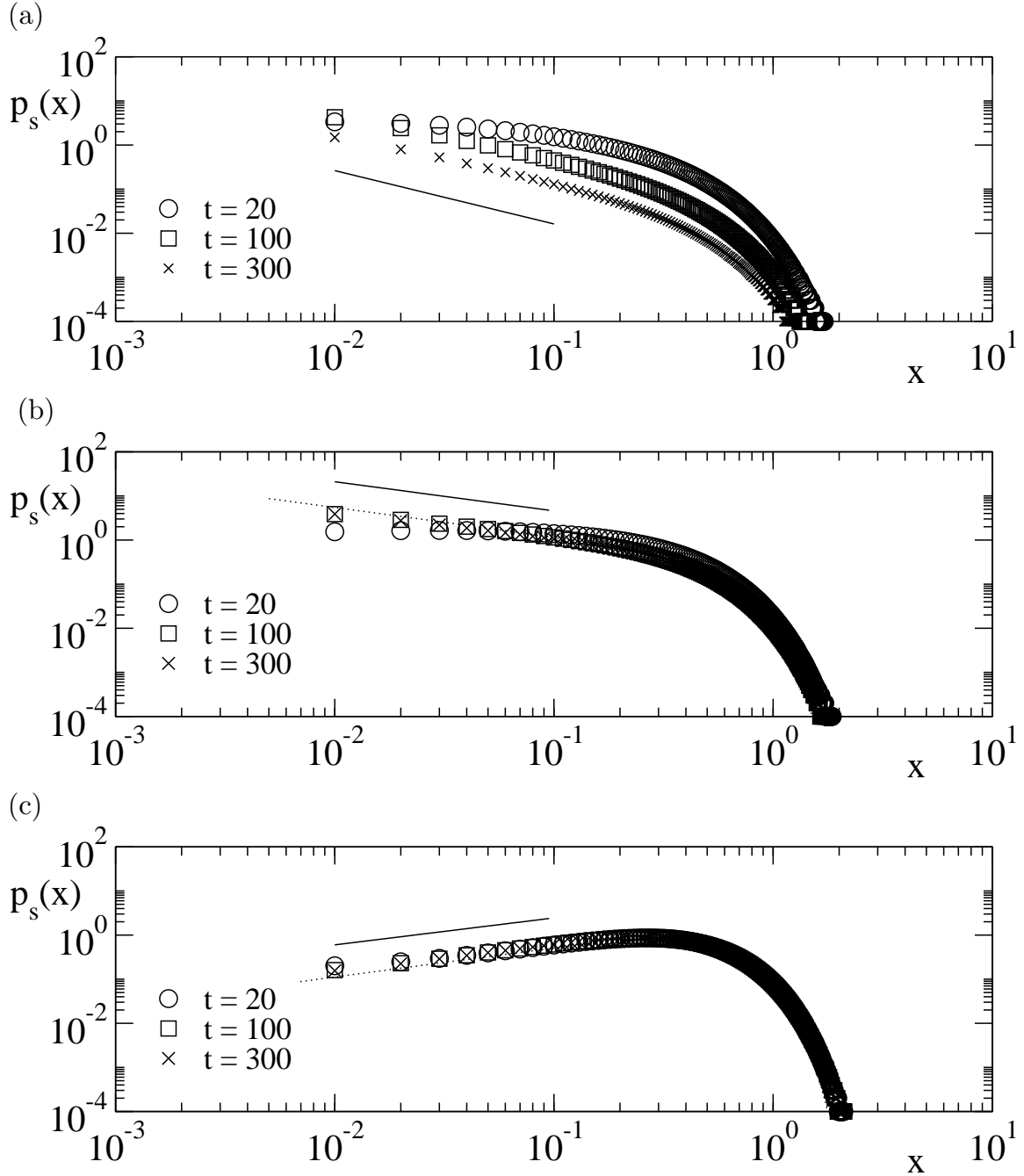}
\caption{Probability distribution function $p(x)$ for $b = 1$, $\tau = 1$ and $D = 0.3$ at the times given, and averaged over $10^{6}$ independent realizations. Values of the control parameter shown are: (a) $a = -1.2$ with $\alpha \simeq -1.21$,  (b)  $a = -1.1$ with $\alpha \simeq -0.66$, and (c) $a = -0.9$ with $\alpha \simeq 0.61$. The distributions in (a) show a clear transient, whereas those in (b) and (c) are stationary. The solid line shows the power law at small $x$; the domain covered by the line indicated the range of data that were used to 
estimate $\alpha$, and is placed above or below the curves for clarity. The dashed line is our approximate determination of $p(x)$ in the limit of small $\tau$.}
\label{fig:stationary_distribution}
\end{figure}

\begin{figure}[h]
\centering\includegraphics[width=6in]{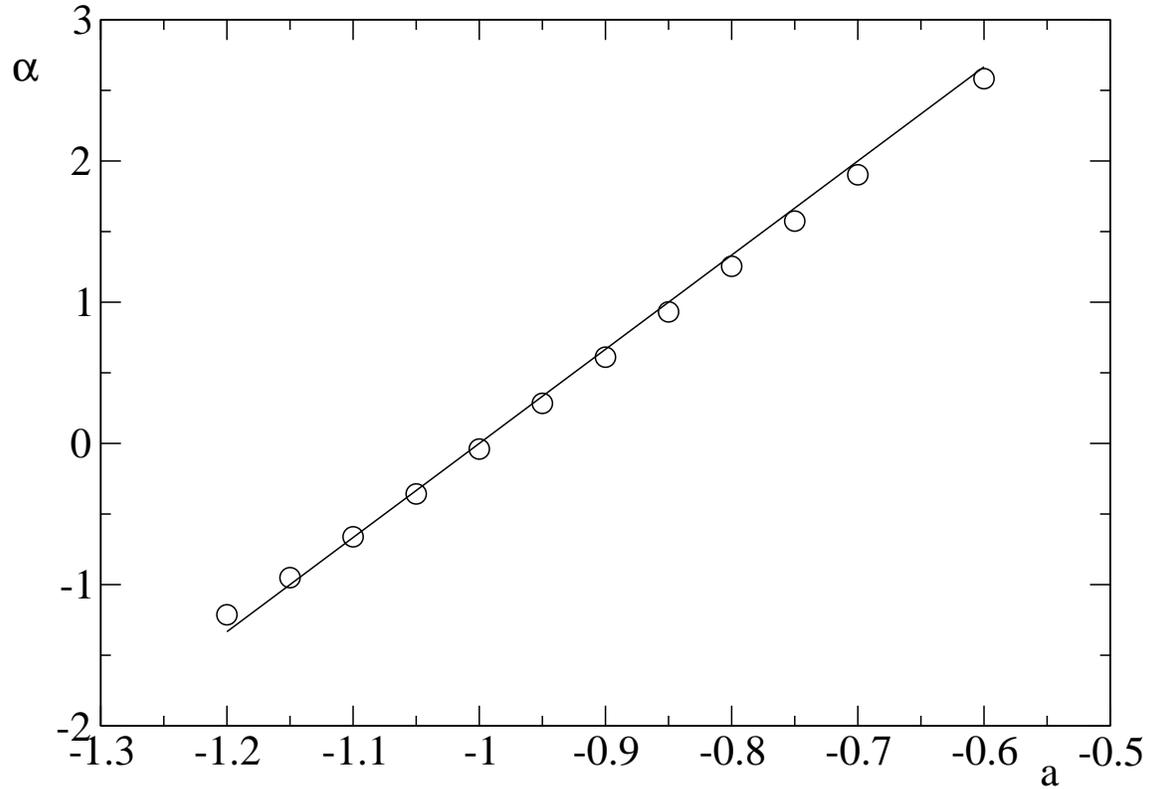}
\vspace{2.0cm}
\caption{Results of a power law fit to $p(x)$ for small $x$. The fitted value of the exponent $\alpha$ is shown as a function of the control parameter $a$. We define the bifurcation threshold for the stochastic problem when $\alpha = -1$, or $a_{c} \simeq -1.18$ for this parameter set ($b = 1$, $\tau = 1$, and $D = 0.3$). The solid line follows from our approximate determination of $p(x)$ in the limit of small $\tau$. There are no adjustable parameters.}
\label{fig:alpha_a}
\end{figure}

Our results for the stationary distribution function $p(x)$ are shown in Fig. (\ref{fig:stationary_distribution}). For $a < a_{c}$, we expect $p(x) = \delta(x)$. We observe instead a very long lasting transient with $p(x)$ approximately characterized by a power law distribution, with an effective exponent $\alpha < -1$ (leading to a non normalizable distribution). The amplitude of the point $p(x \simeq 0)$ (not shown in the figure) grows with time, signaling the build up of the delta function at the origin. Because of overall normalization, growth at $x = 0$ implies a decaying amplitude for finite $x$, as shown in the figure. For $a > a_{c}$, we do obtain a time independent power law distribution $p(x)$ with exponent $-1 < \alpha < 0$. This distribution is normalizable, and represents the stationary distribution above threshold. We finally show $p(x)$ in the range of $a$ for which the distribution is bimodal. The function $p(x)$ around the most likely value is approximately constant over time, but we still observe some transients in the region around $x=0$. Figure (\ref{fig:alpha_a}) shows our results for the exponent of a power law fit to $p(x)$ at small $x$, as a function of the value of the control parameter $a$. We observe a smooth variation of the exponent $\alpha$ with $a$ that allows a convenient determination of $a_{c}$, the value of $a$ for which $\alpha = -1$. This is the method that we have used to determine the bifurcation threshold in all the results presented below.

\begin{figure}[h]
\centering\includegraphics[width=6in]{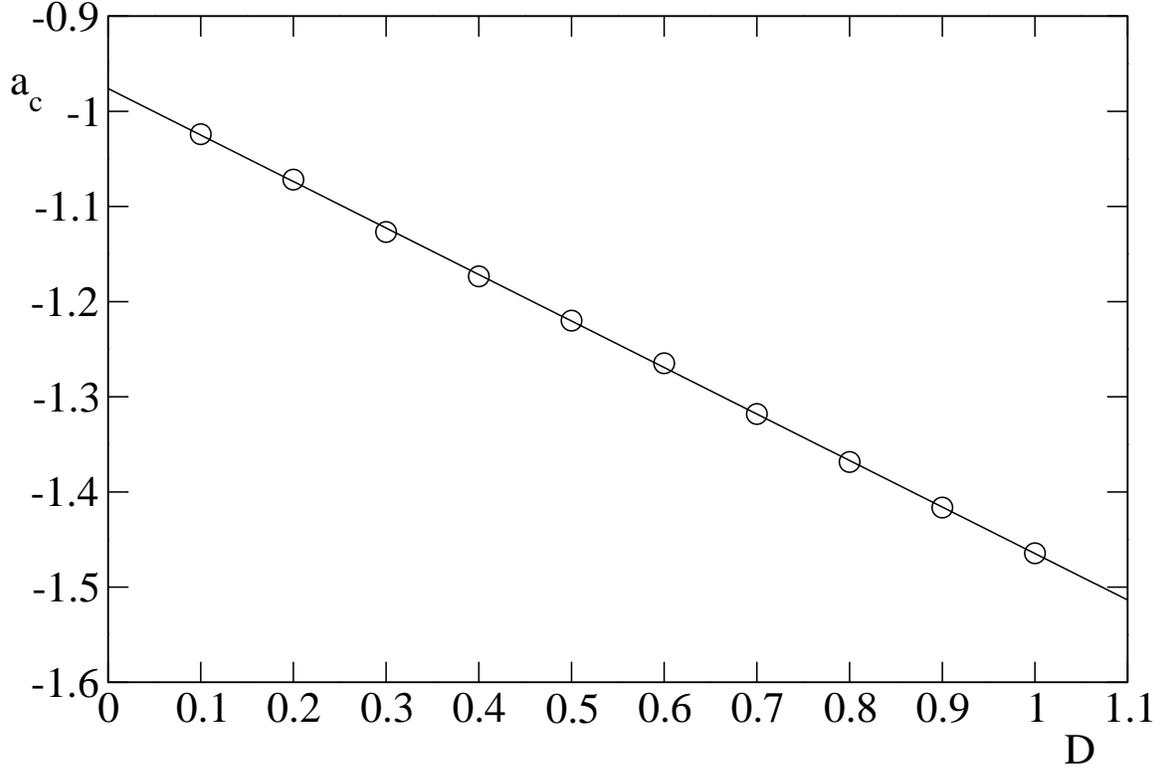}
\vspace{1.0cm}
\caption{Bifurcation threshold $a_{c}$ from $p(x)$ as a function of noise intensity $D$ for $b = 1$ and $\tau = 1$. Time averages used for the determination of $p(x)$ are in $t \in (300,350)$, and $1 \times 10^{6}$ independent realizations have been considered. The line in the figure is the prediction from our approximate determination of the stationary probability distribution function. There are no adjustable parameters.}
\label{fig:noise_shift}
\end{figure}

We summarize our results for the bifurcation diagram of Eq. (\ref{eq:StochmultDDnonlin_ch3}) in Fig. (\ref{fig:bif_diagram}). The analytic results for the threshold without noise ($\xi = 0$) are shown for reference, as well as earlier results for the threshold of $\left\langle x \right\rangle$ of the {\em linearized} equation \cite{re:mackey95}.  Our numerical results do agree with the known threshold for the special point of no delay $b = 0$ given in \cite{re:graham82b}. The figure presents our results for the stochastic bifurcation threshold defined directly from the stationary probability distribution function as discussed above. We conclude that the stochastic threshold is shifted relative to the deterministic threshold except in the special point of no delay ($b = 0$). 
We have also examined in Fig. (\ref{fig:noise_shift}) the dependence of the threshold shift as a function of the noise intensity $D$. In analogy with the case of no delay, we find a linear dependence of $a_{c}$ with $D$.

\subsection{Hopf bifurcation}

\begin{figure}
\centering\includegraphics[width=6in]{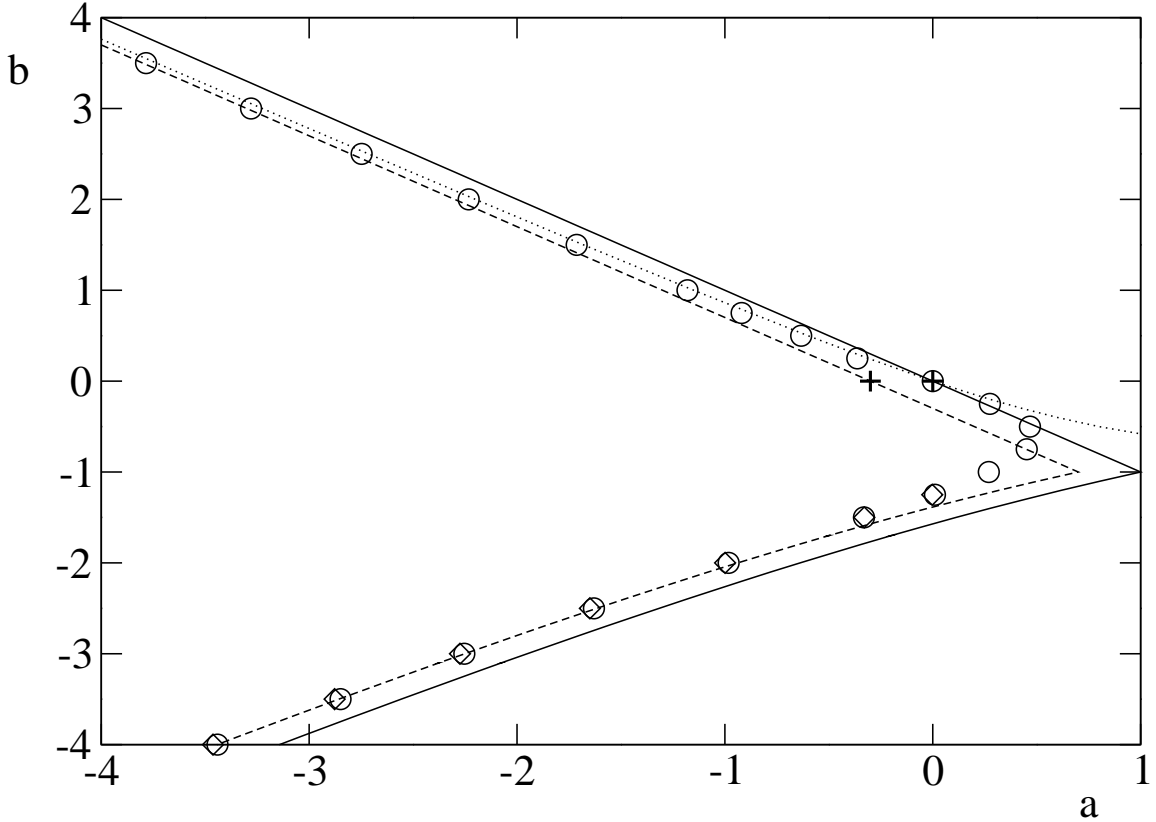}
\vspace{1.0cm}
\caption{Numerically determined bifurcation diagram for Eq. (\ref{eq:StochmultDDnonlin_ch3}) with $\tau = 1$ ($\circ$) defined as the point in the $(a,b)$ plane for which $\alpha = -1$, the exponent of the stationary probability distribution function. For reference, we also show the exact bifurcation diagram for the deterministic equation $\xi = 0$ (solid line), and the exact results for the bifurcation threshold of the first moment $\langle x \rangle$ from the linearized equation \cite{re:mackey95} (dashed line). The two points labeled by $+$ lie on the line $b=0$, and are known results for the case of no delay for (from left to right) $\langle x \rangle$, and for deterministic case of $\xi = 0$. The dotted line is the approximate threshold [Eq. (\ref{eq:thresh_px})], and the labels ($\diamond$) are the numerically calculated Hopf branch from the probability distribution function of the maximum amplitude of the Fourier transform of the trajectories.} 
\label{fig:bif_diagram}
\end{figure}

The calculation of $p(x)$ just shown for the case of a pitchfork bifurcation has been repeated in the vicinity of the deterministic Hopf branch shown in Fig. (\ref{fig:bif_diagram}). In the deterministic case, the bifurcation leads to oscillation. When fluctuations are added, oscillation amplitudes fluctuate as well. We also observe in this range of parameters a sharp bifurcation threshold which is shifted relative to the deterministic Hopf bifurcation. The bifurcation 
threshold is verified with the probability distribution function of the maximum amplitude of the Fourier transform of the trajectories. For each trajectory, we calculate the Fourier transform of the stationary solution over a finite 
window in time, identify its maximum amplitude, and construct an histogram over those maxima along the trajectory and over the ensemble.

\begin{figure}
\includegraphics[width=6in,clip]{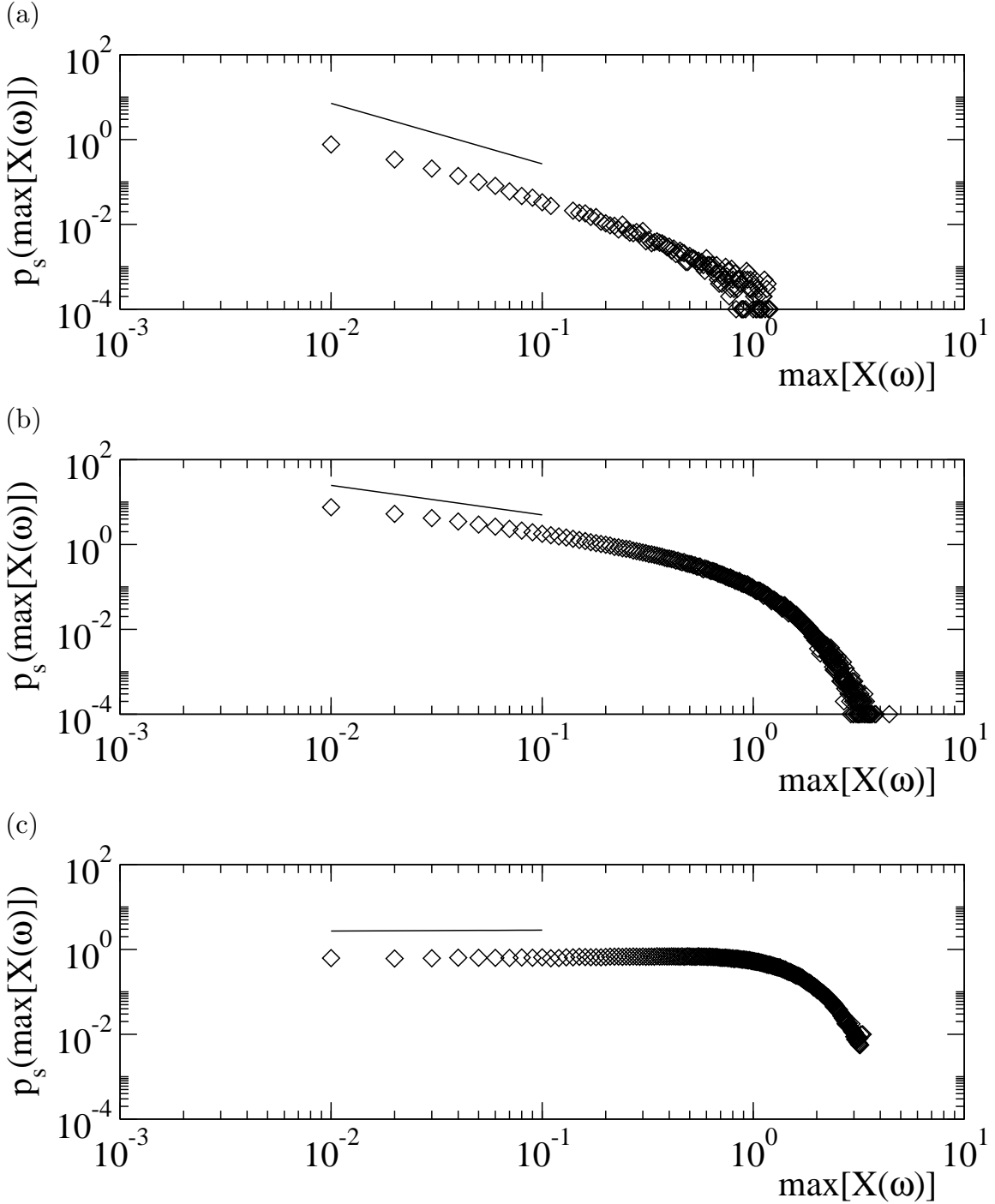}
\caption{Probability distribution function of the maximum amplitude of the Fourier transform $p_{s}(max[X(\omega)])$ for $b=-2$, $\tau=1$ and $D=0.3$ at the times given, and averaged over $10^6$ independent realizations. Values of the control parameter shown are: (a) $a = -1.1$ with $\alpha \simeq -1.43$, (b) $a = -0.9$ with $\alpha \simeq -0.70$, and (c) $a = -0.5$ with $\alpha \simeq 0.02$, as shown by the solid lines, placed above the curves for clarity. The solid lines 
extend over the range used for estimation of $\alpha$. The distributions in (a) show a clear transient, whereas those in (b) and (c) are stationary.}
\label{fig:proghopf_nonlin}
\end{figure}

Our results for the stationary probability distribution function of the maximum amplitude of the Fourier transform are shown in Fig. (\ref{fig:proghopf_nonlin}) for three increasing values of the control parameter $a$. For $a < a_{c}$ we observe an effective power law with exponent $\alpha < -1$ and hence not a physical distribution. As was the case above, this is manifested by a transient distribution that decreases with time. As the value of $a$ is increased, a power law distribution is found with exponent $-1 < \alpha < 0$. The distribution obtained is integrable and stationary. For yet larger values of $a$, the distribution becomes bimodal. 

\begin{figure}
\centering\includegraphics[width=6in]{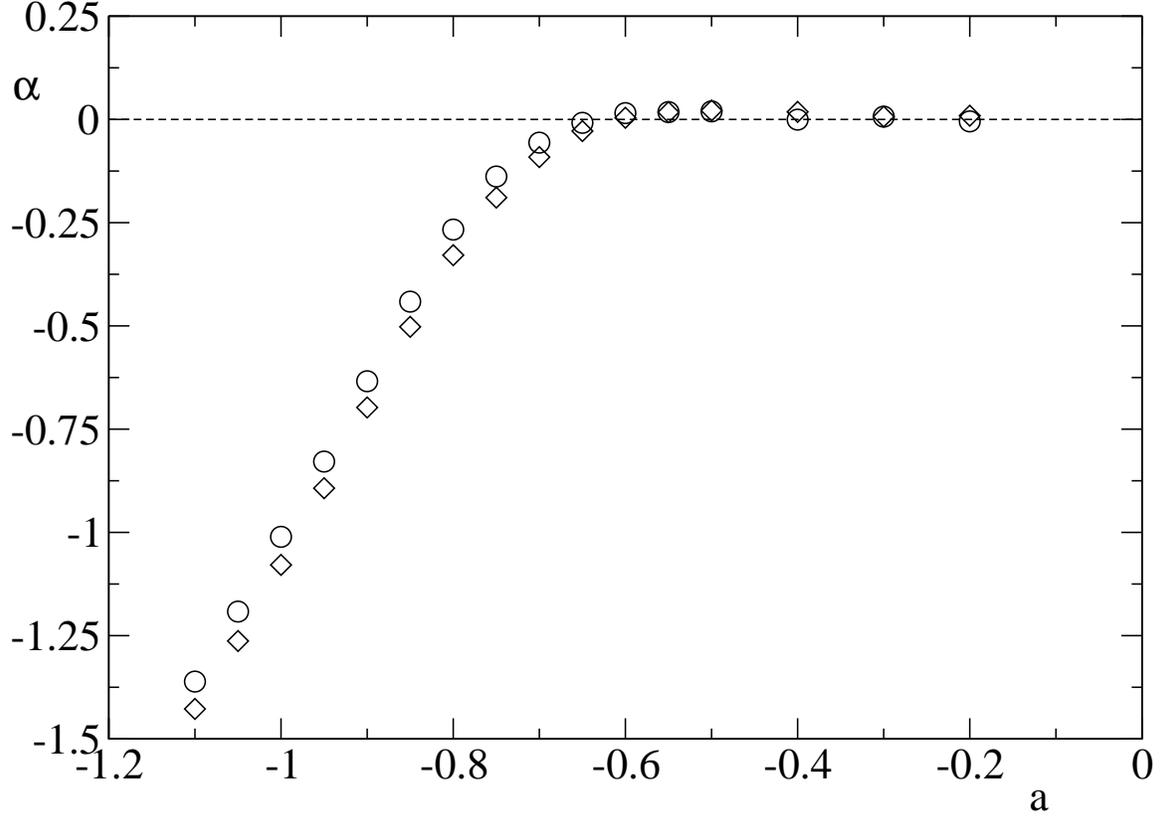}
\vspace{1.0cm}
\caption{Results of a power law fit to $p(x)$ ($\circ$) and to the probability of the maximum amplitude of the Fourier transform $p_{s}(max[X(\omega)])$ ($\diamond$), for small $x$ or small $max[X(\omega)]$. The fitted value of the exponent $\alpha$ is shown as a function of the control parameter $a$. We define the bifurcation threshold for the stochastic problem when $\alpha = -1$, or $a_{c} \simeq -0.999$ ($\circ$), and $a_{c} \simeq -0.981$ ($\diamond$) for this parameter set ($b = -2$, $\tau = 1$, and $D = 0.3$).}
\label{fig:bifu_bm2_1}
\end{figure}

Figure (\ref{fig:bifu_bm2_1}) shows the results of a power law fit to the resulting distributions at small $x$ for a range of values of $a$ of the two probability distribution functions introduced. We have undertaken this analysis for a range of values of $b$, and the resulting Hopf branch of the bifurcation diagram is shown in Fig. (\ref{fig:bif_diagram}). Note that it is also shifted relative to the branch corresponding to the deterministic equation. 

\section{Fokker-Planck equation}

We next turn to an approximate analytic calculation of the stationary distribution function $p(x)$ for Eq. (\ref{eq:StochmultDDnonlin_ch3}). The difficulty in obtaining a closed, analytic expression lies in the non Markovian nature of Eq. (\ref{eq:StochmultDDnonlin_ch3}) and the associated need to find the joint probability distribution $p(x(t),x(t-\tau))$. When $\tau$ is larger than the correlation time of $x$, the derivation is simplified by assuming statistical independence between $x(t)$ and $x(t-\tau)$, or $p(x(t),x(t-\tau)) = p(x(t))p(x(t-\tau))$. This approximation has already been considered in the literature (e.g., ref. \cite{re:bratsun05}). However, the assumption of independence does not hold near a bifurcation since the characteristic correlation time diverges \cite{re:berbert09}. We instead proceed as follows: Define the probability distribution of $x$ as $p(x,t) = \left\langle \delta (x(t)-x) \right\rangle$. By using the properties of the Dirac delta function, and Eq. (\ref{eq:StochmultDDnonlin_ch3}) one finds
\begin{equation}
\frac{\partial}{\partial t} p(x,t) =  - \frac{\partial}{\partial x} \left[\left\langle\delta(x(t)-x)
(ax + bx_{\tau}-x^3) \right\rangle +
\left\langle \delta(x(t)-x)x\xi(t) \right\rangle \right] \; ,
\label{eq:5}
\end{equation}
where we have introduced the dummy variables $x(t) \rightarrow x$ and $x(t-\tau) \rightarrow  x_{\tau}$. The second term in the right hand side of Eq. (\ref{eq:5}) can be written as
\begin{equation}
\left\langle\delta(x(t)-x)x\xi(t)\right\rangle =
x(t)\left\langle\delta(x(t)-x) \xi(t) \right\rangle  \; ,
\label{eq:2eterme_de_5}
\end{equation}
with
\begin{equation}
\left\langle\delta(x(t)-x)\xi(t)\right\rangle = \int^{t'}_0
\left\langle\xi(t)\xi(t')\right\rangle
\left\langle
\frac{\delta  \left[\delta(x(t)-x)\right]}{\delta \xi(t')} \right\rangle dt' =
D\left\langle\frac{\delta \left[\delta(x(t)-x)\right]}{\delta
  \xi(t')}\right\rangle \; ,
\label{eq:2eterme_de_5_2}
\end{equation}
by using the Furutsu-Novikov theorem. By using the properties of the delta function, one can write
\begin{equation}
\left\langle\frac{\delta \left[\delta(x(t)-x)\right]}{\delta
\xi(t')}\right\rangle = \left\langle-\frac{\partial}{\partial x}
\delta(x(t)-x)\frac{\delta x(t)}{\delta
  \xi(t')} \right\rangle \; ,
\end{equation}
and since $ \frac{\delta x(t)}{\delta \xi(t')}=x(t)$ from Eq. (\ref{eq:StochmultDDnonlin_ch3}), we find
\begin{equation}
\left\langle \delta(x(t)-x)\xi(t)\right\rangle = -D \frac{\partial}{\partial x}
x(t)p(x,t)\; ,
\end{equation}
where we have also used
\begin{equation}
\left\langle\delta (x(t)-x)f(x) \right\rangle = \int_{- \infty}^{\infty}
\delta(x(t)-x)f(x) p(x,t) dx = f(x(t)) p(x,t) \; .
\end{equation}
Thus the second term in the right had side of Eq. (\ref{eq:5}) is
\begin{equation}
\begin{split}
-\frac{\partial}{\partial x} \left[ \left\langle\delta(x(t)-x)x(t)\xi(t) \right\rangle \right] &= 
D\frac{\partial}{\partial x} \left[ x(t)\frac{\partial}{\partial x} x(t)p(x,t) \right] \\
&= -D \frac{\partial}{\partial x} \left[x(t)p(x,t)\right] + D\frac{\partial^{2}}{\partial x^{2}} \left[ x^{2}(t)p(x,t) \right] \; .
\end{split}
\label{eq:9}
\end{equation}

Consider now the first term in the right hand side of Eq. (\ref{eq:5}). By using the law of total expectation, 
\begin{equation}
\left\langle \delta(x(t)-x) (ax+bx_{\tau}-x^3) \right\rangle = \int \int \delta(x(t)-x)(ax+bx_{\tau}-x^3) p(x,x_{\tau}) dx dx_{\tau} \; ,
\end{equation}
We anticipate that independence, $p(x,x_{\tau}) = p(x)p(x_{\tau})$, does not hold near the bifurcation threshold. We write instead $p(x,x_{\tau}) = p(x_{\tau}|x)p(x)$, where $p(x_{\tau}|x)$ is the conditional probability of finding $x_{\tau}$ at $t-\tau$ given the information that $x(t)=x$. Then
\begin{eqnarray}
\left\langle \delta(x(t)-x)(ax+bx_{\tau}-x^3)\right\rangle & = & \int \int
\delta(x(t)-x)(ax+bx_{\tau}-x^3) p(x_{\tau}|x)p(x) dx dx_{\tau}\nonumber\\
& = & p(x,t)\int \left[ax(t)+bx_{\tau}-x^3(t)\right]p(x_{\tau}|x(t))dx_{\tau} \nonumber\\
& = & p(x,t)\left[ax(t)-x^3(t) + b\int x_{\tau} p(x_{\tau}|x(t))dx_{\tau} \right]\; .
\label{eq:7}
\end{eqnarray}
The last integral represents the conditional expected value of $x(t-\tau)$
given $x(t)$,
\begin{equation}
\left\langle x_{\tau}|x(t)\right\rangle = \int x_{\tau} p(x_{\tau}|x(t))dx_{\tau} \; .
\label{eq:8}
\end{equation}
Substitution of Eqs. (\ref{eq:9}) and (\ref{eq:7}) into Eq. (\ref{eq:5}) leads 
to the Fokker-Planck equation,
\begin{equation}
\frac{\partial}{\partial t} p(x,t) = -\frac{\partial}{\partial x} \left[ 
(a+D)x(t) - x^{3}(t) + b\langle x_{\tau} | x(t) \rangle  \right]p(x,t) + 
D \frac{\partial^{2}}{\partial x^{2}} \left[ x^{2}(t)p(x,t) \right] \; .
\label{eq:12}
\end{equation}

We have not been able to determine $\left\langle x_{\tau}|x(t) \right\rangle$ analytically. However, it is possible to derive an approximate expression under the assumption that the delay $\tau$ is small \cite{re:frank05}. We write drift 
terms in Eq. (\ref{eq:StochmultDDnonlin_ch3}) in the Ito representation (recall to we were working with the Stratonovich representation), $f(x,x_{\tau}) = (a+D)x - x^{3} + bx_{\tau}$ \cite{re:gardiner85}. Define $\tilde{f}(x) = (a+D)x - x^{3}$, $f^{(0)}(x) = \tilde{f}(x) + Bx$, and $g(x) = x$, so that $f(x,x_{\tau}) = \tilde{f}(x) + Bx_{\tau}$. Furthermore, since we are in the small time delay approximation, one can assume that the zero-th order approximation $p_{st}^{(0)}(x_{\tau},t-\tau|x,t)$ of the stationary conditional probability distribution function $p_{st}(x_{\tau},t-\tau|x,t)$ is Gaussian \cite{re:risken89} and given by,
\begin{equation}
p_{st}^{(0)}(x_{\tau},t-\tau|x,t) = \sqrt{\frac{1}{2\pi\tau g^{2}(x)}} \exp\left(-\frac{\left[x_{\tau}-x-f^{(0)}(x)\tau\right]^{2}}{2\tau g^{2}(x)}\right) \;.
\end{equation}
The approximate drift term of the Fokker-Planck equation is then, 
\begin{equation}
f(x) = \sqrt{\frac{1}{2\pi\tau x^{2}}} \int_{-\infty}^{\infty} dx_{\tau}
f(x,x_{\tau}) \exp \left(-\frac{\left[x_{\tau}-x-f^{(0)}(x)\tau\right]^{2}}{2\tau x^{2}}\right) \;.
\end{equation}
Integrating, one finds, 
\begin{equation}
f(x) = (1+b\tau)f^{(0)}(x) = (1+b\tau)\left[(a+b+D)x - x^{3}\right] \;.
\label{eq:effdrift}
\end{equation}
An approximate expression for the conditional average of $x_{\tau}$ given $x(t)$ is obtained by comparing Eq. (\ref{eq:effdrift}) with Eq. (\ref{eq:7}), 
\begin{equation}
\left\langle x_{\tau}|x(t)\right\rangle = \left[1+\tau(a+b+D)\right] x(t) - \tau x^{3} \; .
\label{eq:condrifttheo}
\end{equation}
Further substitution of Eq. (\ref{eq:condrifttheo}) into Eq. (\ref{eq:12}) results in a closed form for the steady state distribution, 
\begin{equation}
\frac{\partial}{\partial t} p(x,t) = -\frac{1}{(1+b\tau)}\frac{\partial}
{\partial x} \left[ \left\{ \left(a+b+D\right)x(t) - x^{3}(t) \right\} 
p(x,t)\right] + \frac{D}{\left(1+b\tau\right)^{2}} 
\frac{\partial^{2}}{\partial x^{2}} \left[ x^{2}(t)p(x,t) \right] \; ,
\label{eq:11}
\end{equation}
with stationary solution,
\begin{equation}
p(x) = N |x|^{\frac{a+b[1+\tau(a+b+D)]}{D}-1}e^{-\frac{(1+b\tau)}{2D}x^{2}} \; .
\label{eq:po_DDnonlin}
\end{equation}
This stationary distribution is found to agree quite well with our numerical determination of $p(x)$ (Fig. (\ref{fig:stationary_distribution})) around the pitchfork branch, but fails around the Hopf branch. Furthermore, we can now introduce an analytic determination of the bifurcation threshold as the point in which $p(x)$ (Eq. (\ref{eq:po_DDnonlin})) ceases be normalizable:  $\alpha_{c} = \frac{a_{c}+b[1+\tau(a_{c}+b+D)]}{D} - 1 = -1$. The bifurcation threshold is located at
\begin{equation}
a_{c} = -\frac{b[1+\tau(b+D)]}{1+b\tau} .
\label{eq:thresh_px}
\end{equation}
This prediction is also in very good agreement with our numerical results 
for the pitchfork bifurcation branch.

\begin{figure}
\centering\includegraphics[width=6in]{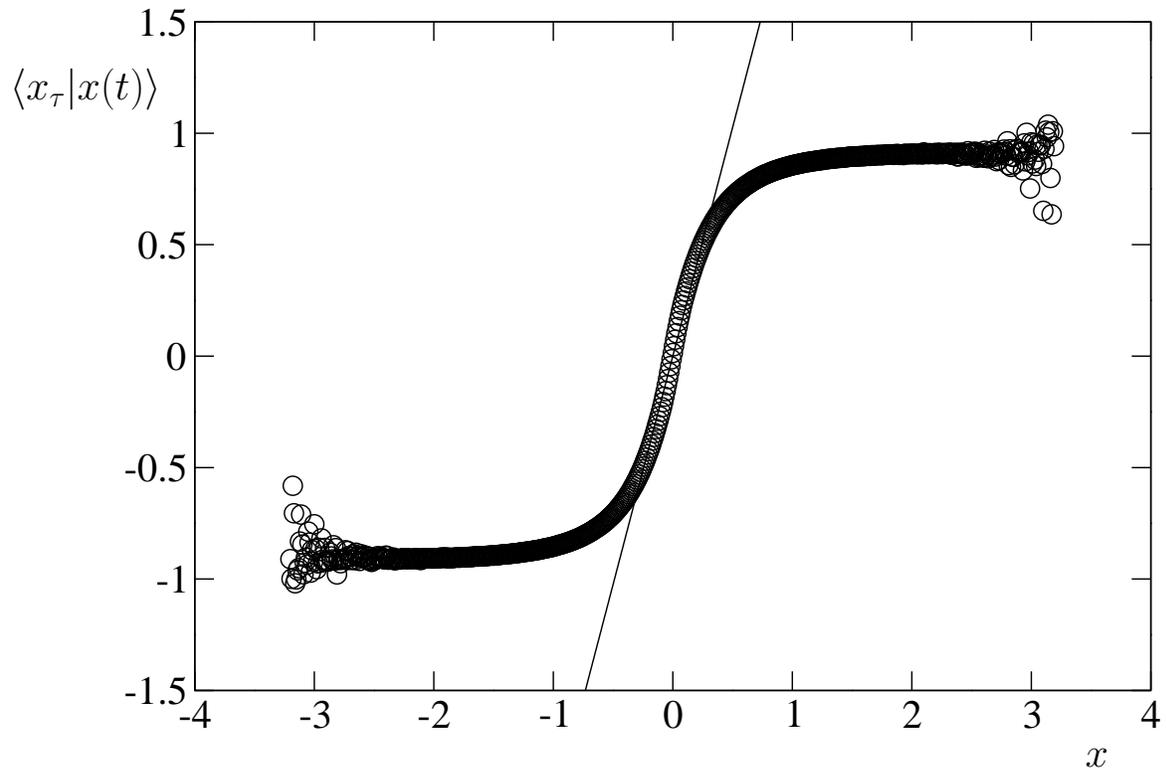}
\vspace{1.0cm}
\caption{Ensemble average of $x(t-\tau)$ given $x(t)$, $\left\langle x_{\tau} | x(t) \right\rangle$. We have set $a = 1,b = -0.25, \tau =1$ and $D = 0.3$ The average is computed over $10^{6}$ realizations for a time interval $t \in (300,350)$. The dashed line is $\left\langle x_{\tau} | x(t) \right\rangle = (1 + \tau(a + b + D)) x(t)$, the approximate analytic result in the limit of small $\tau$.}
\label{fig:CAD}
\end{figure}

We have also verified our results for the conditional average with a
direct numerical integration of the model equation. We show our results 
for $\left\langle
x_{\tau}|x(t) \right\rangle$ as a function of $x(t)$ in Fig. (\ref{fig:CAD}). 
Equation (\ref{eq:StochmultDDnonlin_ch3}) is integrated numerically until 
it reaches a statistical stationary state. For every value of the 
dynamical variable $x$, the average of its value at time $t-\tau$ 
is collected for $1 \times 10^{6}$ independent trajectories in a time 
window $t \in (290,300)$. We note that for large values of $x$, the results 
become less accurate because of fewer data points in this region. This 
analysis has been repeated for a range values of $a$, $b$, and $D$. 
As shown in the figure, we observe good agreement between the numerical 
results and Eq. (\ref{eq:condrifttheo}) in the region around $x = 0$. 

In summary, we find that the bifurcation point in our stochastic differential equation with delay remains sharp. This is not a straightforward observation since the delay term in Eq. (\ref{eq:StochmultDDnonlin_ch3}) could effectively act as an additive source of noise, and lead to an imperfect bifurcation instead. This does not appear to be the case. We also note that long transients can be expected below the bifurcation threshold, as all trajectories eventually decay to the trivial solution $x = 0$.  Second, we observe that different moments of $x$ obtained from the {\em linearized} equation bifurcate at different values of the control parameter. When a saturating nonlinearity is introduced into the model, all moments bifurcate at $a_{c}$. By defining the bifurcation point in the stochastic case as that in which the power law form of $p(x)$ becomes normalizable ($\alpha = -1$), we show that the bifurcation threshold $a_{c}$ is shifted relative to that of the underlying deterministic equations. The shift goes to zero as $b \rightarrow 0$, and otherwise it scales linearly with the noise intensity $D$. We have also derived and approximate expression for the stationary distribution function $p(x)$ in the limit of small delay time $\tau$. The threshold location obtained from this approximation agrees well with our numerical determination even when $\tau = 1$.

\section*{Acknowledgments}
This research has been supported by NSERC Canada. MG acknowledges funding by the FQRNT. Computational resources have been provided by CLUMEQ.


\newpage

\appendix
\section{Algorithm for stochastic differential equations with delay} \label{se:appA}

We summarize in this appendix the numerical algorithm developed to integrate Eq. (\ref{eq:StochmultDDnonlin_ch3}). Formal integration yields,
\begin{equation}
x(t + \Delta t) = x(t) + \int^{t+\Delta t}_t ax(t')dt' -\int^{t'}_t
x^3(t')dt' + \int^{t+\Delta t}_{t} x(t')\xi(t')dt' + \int^{t+\Delta
  t}_t bx(t' - \tau )dt'.
\label{eq:algo1_2}
\end{equation}
with
\begin{equation}
x(t') = x(t) + \int^{t'}_t
ax(t'')dt'' -\int^{t'}_t x^3(t'')dt'' + \int^{t'}_t x(t'')\xi(t'')dt''
+ \int^{t'}_t bx(t'' - \tau )dt'' .
\end{equation}
We now approximate $x(t'') \approx x(t)$ and $ x(t''-\tau) \approx x(t-\tau)$ so that to first order in $(t'-t)$ we have,
\begin{equation}
x(t') = x(t)+ \left[ ax(t)+bx(t-\tau)-x^3(t) \right](t'-t) +
x(t)\int^{t'}_t \xi(t'')dt'' .
\label{eq:algo40_2}
\end{equation}
The nonlinear term $x^3(t')$ also must be expanded around $x(t)$
\begin{equation}
x^3(t') \approx x(t)^{3} + 3x^2(t)(x(t')-x(t)) \; .
\label{eq:x3}
\end{equation}
We substitute Eqs. (\ref{eq:algo40_2}) and (\ref{eq:x3}) into equation Eq.
(\ref{eq:algo1_2}) and find,
\begin{eqnarray}
x(t + \Delta t) - x(t) & = & ax(t)\Delta t + \left[a^{2}x(t) + a b x(t
- \tau ) - ax^3(t)\right] \frac{\Delta t^2}{2} + ax(t) \int^{t+\Delta
  t}_t \int^{t'}_t \xi(t'')dt''dt' \nonumber\\
& + & x(t) \int^{t+\Delta t}_t \xi(t')dt' + \left[ax(t) + bx(t - \tau
)- x^3(t)\right] \int^{t+\Delta t}_t (t' - t)\xi(t')dt'\nonumber\\
&+& x(t) \int^{t+\Delta t}_t \xi(t') \int^{t'}_t \xi(t'')dt''dt' + bx(t -
\tau)\Delta t \nonumber\\
& + & \left[b a x(t - \tau ) + b^2 x(t - 2\tau )-b x^3(t-\tau)\right]
\frac{\Delta t^2}{2}  \nonumber\\
& + & bx(t - \tau ) \int^{t+\Delta t}_t
\int^{t'-\tau}_{t-\tau}\xi(t'')dt''dt' - x^3(t)\Delta t \nonumber\\
&-&3x^2(t)\left[ax(t) + bx(t - \tau)-x^3(t)\right]\frac{\Delta
  t^2}{2}-3 x^3(t) \int^{t+\Delta t}_t \int^{t'}_t \xi(t'')dt''dt'
\nonumber .
\end{eqnarray}

In order to calculate the integrals containing the random process
$\xi(t)$, we define
\begin{equation}
\int^{t+\Delta t}_t \xi(t')dt' = G_1(t,\Delta t) \; ,
\label{eq:algo49}
\end{equation}
\begin{equation}
\int^{t+\Delta t}_t \int^{t'}_t \xi(t'')dt''dt' = G_2(t,\Delta t) \; .
\label{eq:algo48}
\end{equation}
If $\xi(t)$ is a Gaussian process of zero mean, $G_1$ and $G_2$ are also Gaussian variables of zero mean, and correlations
\begin{equation}
\left\langle G_1^2 \right\rangle = 2D\Delta t \; ,
\label{eq:algo53}
\end{equation}
\begin{equation}
\left\langle G_2^2 \right\rangle = \frac{2D}{3} \Delta t^3 \; ,
\label{eq:algo54}
\end{equation}
\begin{equation}
\left\langle G_1 G_2 \right\rangle = D\Delta t^2 \; .
\label{eq:algo55}
\end{equation}

The three remaining integrals can be expressed in terms of $G_{1}$ and
$G_{2}$ as
\begin{equation}
\int^{t+\Delta }_t (t' - t)\xi(t')dt' = G_1(t,\Delta t)\Delta t -
G_2(t,\Delta t)  \; ,
\label{eq:algo50}
\end{equation}
\begin{equation}
\int^{t+\Delta }_t \xi(t') \int^{t'}_t \xi(t'')dt''dt' = \frac{1}{2}
(G_1(t,\Delta t))^2  \; ,
\label{eq:algo51}
\end{equation}
\begin{equation}
\int^{t+\Delta t}_t \int^{t'-\tau}_{t-\tau} \xi(t'')dt''dt' = G_2(t -
\tau,\Delta t) \; .
\label{eq:algo52}
\end{equation}
The Gaussian variables $G_1$ and $G_2$ can be simulated with two Gaussian random variables, $\Psi_1$ and $\Psi_2$, of zero mean and
variance one:
\begin{equation}
G_1(t, \Delta t) = \sqrt{2D \Delta t} \Psi_1(t) \; ,
\label{eq:algo56}
\end{equation}
\begin{equation}
G_2(t, \Delta t) = \sqrt{\frac{2D}{3}\Delta
  t^3}\left(\frac{\sqrt{3}}{2}\Psi_1(t) + \frac{1}{2}\Psi_2(t)\right)
\; .
\label{eq:algo57}
\end{equation}

Combining all our results, we write the iteration of our algorithm,
\begin{eqnarray}
x(t + \Delta t) & = & x(t)\left(1+a\Delta t + a^2\frac{\Delta t^2}{2}+
(1 + a\Delta t)G_1(t,\Delta t) + \frac{1}{2}(G_1(t,\Delta t))^2\right)
\nonumber\\
& + & x(t - \tau) \left(b\Delta t + 2ab \frac{\Delta t^2}{2} + b\Delta
t G_1(t,\Delta t) - b G_2(t,\Delta t) + b G_2(t - \tau,\Delta
t)\right)
\nonumber\\
& + & x(t - 2\tau)\left(b^2\frac{\Delta t^2}{2}\right) + x^3(t)\left(
-2a\Delta t^2 - (G_1(t,\Delta t)+1)\Delta t -2G_2\right)\nonumber\\
& - & x^2(t)x(t-\tau)\left(\frac{3 b \Delta t^2}{2}\right) -
x^3(t-\tau)\left(\frac{b
  \Delta t^2}{2}\right) + x^5(t)\left(\frac{3\Delta t^2}{2}\right) \; .
\end{eqnarray}

\newpage

\begin{center}
FIGURES
\end{center}

\newpage

\newpage

\newpage

\newpage

\newpage

\newpage

\end{document}